\documentclass[12pt, notitlepage]{article}
\pdfoutput=1
\usepackage{float}
\usepackage{subfig}
\usepackage{graphicx}
\usepackage[T1]{fontenc}
\usepackage[utf8]{inputenc}
\usepackage{authblk}
\usepackage{amsmath,bm}
\usepackage[left=3cm,right=3cm,top=2cm,bottom=2cm]{geometry}
\renewcommand\footnotemark{}
\begin{document}


\title{Evidence for Asymptotic Safety from Dimensional Reduction in Causal Dynamical Triangulations}
\author{D.N.~Coumbe}     
\author{J. Jurkiewicz}
\affil{\emph{Faculty of Physics, Astronomy and Applied Computer Science, Jagiellonian University, ul. prof. Stanislawa Lojasiewicza 11, Krakow, PL 30-348}\footnote{E-mail: daniel.coumbe@uj.edu.pl, jerzy.jurkiewicz@uj.edu.pl}}

\date{\today}          
\maketitle


\begin{abstract}

We calculate the spectral dimension for a nonperturbative lattice approach to quantum gravity, known as causal dynamical triangulations (CDT), showing that the dimension of spacetime smoothly decreases from $\sim4$ on large distance scales to $\sim3/2$ on small distance scales. This novel result may provide a possible resolution to a long-standing argument against the asymptotic safety scenario. A method for determining the relative lattice spacing within the physical phase of the CDT parameter space is also outlined, which might prove useful when studying renormalization group flow in models of lattice quantum gravity. 

\end{abstract}


\begin{section}{Introduction}

Three of the four fundamental interactions of nature have been successfully quantised, the notable exception being gravity. The central difficulty in formulating a theory of quantum gravity is that the computational techniques applied so successfully to the other forces do not give consistent results when applied to quantum general relativity. The origin of this incongruity stems from the fact that gravity is distinguished from the other fundamental interactions of nature by its dimensionful coupling constant $G_{N}$. In $d$-dimensional spacetime Newton's gravitational coupling has a mass dimension of $[G_{N}]=2-d$, meaning that in the case of 4-dimensional spacetime higher-order loop corrections generate a divergent number of counterterms of ever increasing dimension. One can clearly see this from the perturbative quantum field theoretic treatment of gravity in $d$-dimensional space, showing that momentum $p$ scales with loop order $L$ as

\begin{equation}
\int p^{A-[G_{N}]L}dp,
\label{divmom}
\end{equation}

\noindent where $A$ is a process dependent quantity that is independent of $L$ \cite{Weinberg79}. Equation (\ref{divmom}) is clearly divergent for $[G_{N}]<0$, because the integral will grow without bound as the loop-order $L$ increases in the perturbative expansion \cite{Weinberg79}. Interestingly, Eq. (\ref{divmom}) is divergence free for $d\leq2$, meaning that gravity as a perturbative quantum field theory can be renormalizable by power counting if the dimension of spacetime is equal to, or smaller than, two. This raises the exciting possibility that spacetime could act as its own ultraviolet regulator via the mechanism of dynamical dimensional reduction, possibly yielding a finite and predictive theory of quantum gravity. 

Remarkably, a number of seemingly independent approaches to quantum gravity have reported that the dimension of spacetime exhibits a scale dependence. Causal dynamical triangulations (CDT) \cite{Ambjorn:2005db}, exact renormalization group methods \cite{Lauscher:2005qz}, Ho{\v r}ava-Lifshitz gravity \cite{Horava:2009if}, loop quantum gravity \cite{Modesto:2008jz}, and string theory \cite{Atick:1988si,Calcagni:2013eua} all provide evidence that the dimension of spacetime appears to reduce from approximately four on macroscopic scales to approximately two on microscopic scales. Individually these results do not constitute substantial evidence in support of dimensional reduction; collectively, however, they form a compelling argument that demands further attention. 

One of the original formulations of lattice gravity is Euclidean dynamical triangulations (EDT) \cite{Ambjorn:1991pq,Catterall:1994pg}, which defines a spacetime of locally flat $n$-simplices of fixed edge length, where a $n$-simplex is the $n$-dimensional analogue of a triangle. However, the original EDT model quickly ran into significant problems. The parameter space of couplings contained just two phases, neither of which resembled 4-dimensional semi-classical general relativity, and the two phases were separated by a first order critical point, making it unlikely that one could take a continuum limit \cite{Bialas:1996wu,deBakker:1996zx}. In response to these problems a causality condition was added, giving rise to the method of causal dynamical triangulations (CDT) \cite{Ambjorn:1998xu}.

In close analogy to the sum over all possible paths in Feynman's path integral approach to quantum mechanics, CDT is an attempt to construct a nonperturbative theory of quantum gravity via a sum over different spacetime geometries. In CDT, such spacetime geometries are defined by locally flat $n$-dimensional simplices that are glued together along their $\left(n-2\right)$-dimensional faces, forming a $n$-dimensional simplicial manifold. A key ingredient of CDT is the introduction of a causality condition, in which one distinguishes between space-like and time-like links on the lattice. In this way one can define a foliation of the lattice into space-like hypersurfaces, each with the same fixed topology. Only geometries that can be foliated in this way are included in the ensemble of triangulations that define the path integral measure. 

The introduction of the causality condition in the CDT approach to quantum gravity has produced a number of promising results, in contrast to the original EDT version. A four-dimensional de Sitter like phase was shown to emerge within the parameter space of CDT \cite{Ambjorn:2007jv}, and the likely identification of a second-order phase transition line suggests the exciting possibility that the theory may have a well defined continuum limit \cite{Ambjorn:2011cg}. Another key result is that within the de Sitter-like phase of CDT the dimension appears to be scale dependent, dynamically reducing from approximately four on large scales to approximately two on small scales \cite{Ambjorn:2005db}. Since a scale dependent dimension may have important implications for the renormalizability of quantum gravity it forms the central focus of this work.

At first glance one might think that performing a weighted sum over geometries constructed by gluing together $n$-dimensional building blocks will always result in a $n$-dimensional geometry, however, this is not necessarily the case. For dynamical triangulations the dynamics is contained in the connectivity of the $n$-simplices, where the geometry is updated by a set of local update moves \cite{Ambjorn05}. These local update moves can result in the deletion or insertion of vertices within simplices, and so it is possible to obtain a geometric structure that has self-similar properties at different scales; meaning the geometry can be a fractal. A fractal geometry admits non-integer dimensions, so recovering $n$-dimensional space from $n$-dimensional building blocks is a non-trivial test of the theory; a test that CDT has passed by demonstrating that a four-dimensional geometry emerges on large scales \cite{Ambjorn:2005db}. The CDT approach to quantum gravity allows the fractal dimension of the ensemble of triangulations to be computed numerically, typically this is done by computing the Hausdorff dimension and the spectral dimension.

The Hausdorff dimension \cite{Hausdorff:1919vt} generalises the concept of dimension to non-integer values, and can be defined by considering how the volume of a sphere with topological dimension $D_{T}$ scales with radius $r$ in the limit $r\rightarrow0$,

\begin{equation}
D_{H}=\lim_{r \to 0} \frac{\rm{ln} \left(V\left(r\right)\right)}{\rm{ln} \left(r\right)}.
\label{spec0}
\end{equation}

The spectral dimension, on the other hand, is related to the probability of return $P_{r}\left(\sigma\right)$ for a random walk over the ensemble of triangulations after $\sigma$ diffusion steps. One can derive the spectral dimension (following Refs. \cite{Ambjorn:2005db,Benedetti:2009ge}) starting from the $d$-dimensional diffusion equation,

\begin{equation}
\frac{\partial}{\partial\sigma}K_{g}\left(\zeta_{0},\zeta,\sigma\right)-g^{\mu\nu}\bigtriangledown_{\mu}\bigtriangledown_{\nu}K_{g}\left(\zeta_{0},\zeta,\sigma\right)=0,
\label{diffusion1}
\end{equation}

\noindent where $K_{g}$ is known as the heat kernel describing the probability density of diffusion from $\zeta_{0}$ to $\zeta$ in a fictitious diffusion time $\sigma$. $\bigtriangledown$ is the covariant derivative of the metric $g_{\mu\nu}$. The diffusion process is taken over a $d$-dimensional closed Riemannian manifold $M$ with a smooth metric $g_{\mu\nu}\left(\zeta\right)$.

In the case of infinitely flat Euclidean space, Eq. (\ref{diffusion1}) has the simple solution,

\begin{equation}
K_{g}\left(\zeta_{0},\zeta,\sigma\right)=\frac   {\rm{exp}\left(-d_{g}^{2}\left(\zeta,\zeta_{0}\right)/4\sigma\right)}  {\left(4\pi\sigma\right)^{d/2}},
\end{equation}

\noindent where $d_{g}^{2}\left(\zeta,\zeta_{0}\right)$ is the geodesic distance between $\zeta$ and $\zeta_{0}$.

The quantity that is measured in the numerical simulations is the probability $P_{r}\left(\sigma\right)$ that the diffusion process will return to a randomly chosen origin after $\sigma$ diffusion steps over the spacetime volume $V=\int d^{d}\zeta\sqrt{\rm{det}\left(g\left(\zeta\right)\right)}$,

\begin{equation}
P_{r}\left(\sigma\right)=\frac{1}{V}\int d^{d}\zeta\sqrt{\rm{det}\left(g\left(\zeta\right)\right)} K_{g}\left(\zeta,\zeta,\sigma\right).
\end{equation}
 
\noindent The probability of return to the origin in asymptotically flat space is given by,

\begin{equation}
P_{r}\left(\sigma\right)=\frac{1}{\sigma^{d/2}},
\end{equation}

\noindent and so we can extract the spectral dimension $D_{S}$ by taking the logarithmic derivative with respect to the diffusion time, giving

\begin{equation}
D_{S}=-2\frac{d\rm{log}\langle P_{r}\left(\sigma\right)\rangle}{d\rm{log}\sigma}.
\label{SpecDef}
\end{equation}

Equation (\ref{SpecDef}) is strictly only valid for an infinitely flat Euclidean space. However, one can still use this definition of the spectral dimension to compute the fractal dimension of a curved, or finite volume, by factoring in the appropriate corrections for large diffusion times $\sigma$. Specifically, the probability that the random walk will return to the origin approaches unity as the ratio of the volume and the diffusion time approaches zero, i.e. when the diffusion time is much greater than the volume. The mathematical explanation for this is that the zero mode of the Laplacian $-\bigtriangleup_{g}$, which determines the behaviour of $P_{r}\left(\sigma\right)$ via its eigenvalues $\lambda_{n}$, will dominate the diffusion in this region, causing $P_{r}\left(\sigma\right)\rightarrow1/N_{4}$ for $\sigma\gg N_{4}^{2/D_{S}}$ \cite{Ambjorn:2005db}. One can therefore factor in the appropriate finite volume corrections by omitting values of $D_{S}\left(\sigma\right)$ for which $\sigma\gg N_{4}^{2/D_{S}}$.\interfootnotelinepenalty=10000 \footnote{\scriptsize The curvature of the space on which the diffusion process occurs should also be corrected for due to the fact that it will change the probability that the diffusion process will return to the origin \cite{Ambjorn:2005db}. Curvature corrections are not estimated in this work.} The spectral dimension allows one to probe the geometry of spacetime over varying distance scales. The Hausdorff and spectral dimensions coincide with the standard measure of the dimension, the topological dimension, when the manifold is non-fractal. 

\end{section}

\begin{section}{Asymptotic Safety}

As first suggested by Weinberg \cite{Weinberg79}, the concept of the renormalizability of gravity might be generalised to include the nonperturbative regime via the asymptotic safety scenario. In this scenario gravity would be nonperturbatively renormalizable if a finite number of relevant couplings end on an ultraviolet fixed point (UVFP). In a lattice theory of gravity, such as CDT, an UVFP would appear as a second order critical point, the approach to which would define a continuum limit. 

However, there exists an argument due to Banks \cite{Banks:2010tj} (see also Shomer \cite{Shomer:2007vq}) against the possibility of asymptotic safety. The argument compares the density of states at high energies expected for a theory of gravity to that of a conformal field theory. Since a renormalizable quantum field theory is a perturbation of a conformal field theory by relevant operators, a renormalizable field theory must have the same high energy asymptotic density of states as a conformal field theory. It follows from dimensional analysis, and the extensive scaling of the quantities considered, and the fact that a finite temperature conformal field theory has no dimensionful scales other than the temperature, that the entropy S and energy E scale as

\begin{equation}
S\sim\left(RT\right)^{d-1}, E\sim R^{d-1}T^{d}
\label{eq:Entropy1}
\end{equation}

\noindent where $R$ is the radius of the spatial volume under consideration and $T$ is the temperature. It follows that the entropy of a renormalizable theory must scale as\interfootnotelinepenalty=10000 \footnote{\scriptsize See Ref. \cite{Falls:2012nd} for a critique of the reasoning that leads to this scaling.}

\begin{equation}
S\sim E^{\frac{d-1}{d}}.
\label{eq:Entropy2}
\end{equation}

For gravity, however, one expects that the high energy spectrum will be dominated by black holes.\interfootnotelinepenalty=10000 \footnote{\scriptsize Although this assumption has been questioned by Percacci and Vacca \cite{Percacci:2010af}, among others.} The $d$-dimensional Schwarzschild solution in asymptotically flat spacetime has a black hole with event horizon of radius $r^{d-3}\sim G_{N}M$, where $M$ is the mass of the black hole.\interfootnotelinepenalty=10000 \footnote{\scriptsize Asymptotically safe black holes are actually Schwarzschild-de Sitter black holes whose entropy is given by the Cardy-Verlinde formula, which may itself resolve the apparent contradiction between black hole entropy and asymptotic safety \cite{Koch:2013owa}.} The Bekenstein-Hawking area law tells us that $S\sim r^{d-2}$, so that

\begin{equation}
S\sim E^{\frac{d-2}{d-3}}.
\label{eq:Entropy3}
\end{equation}

\noindent This scaling disagrees with that of Eq. (\ref{eq:Entropy2}). Assuming the argument leading to Eq. (\ref{eq:Entropy3}) is valid then one is led to conclude that gravity cannot be formulated as a renormalizable quantum field theory. This is a potentially serious obstacle for asymptotic safety, a possible resolution of which is provided in the following section.


\end{section}


\begin{section}{Measurements of the spectral dimension in CDT}

\begin{figure}[H]
  \centering
  \includegraphics[width=1.0\linewidth,natwidth=610,natheight=642]{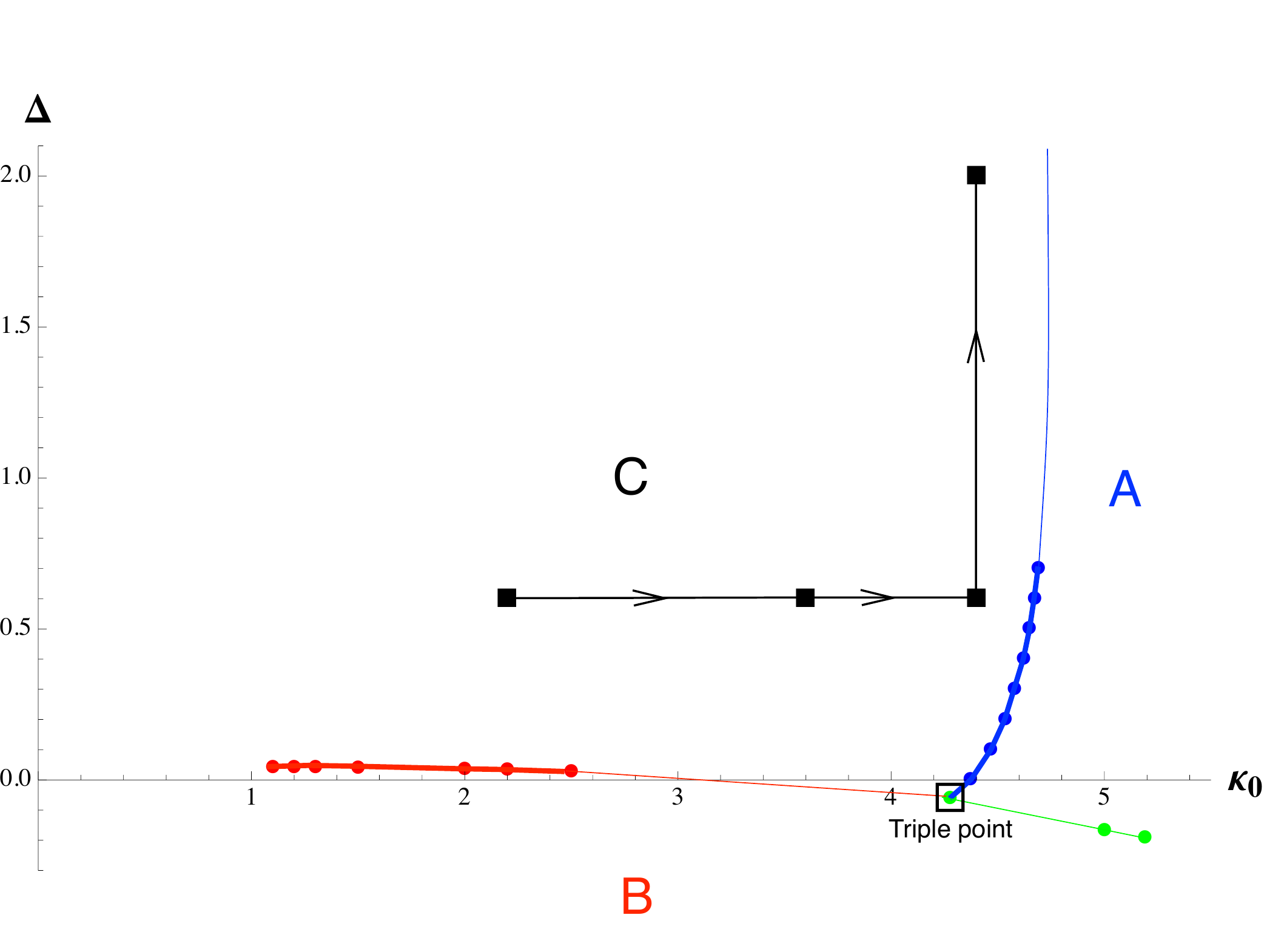}
  \caption{\small A schematic representation of the phase diagram of 4-dimensional CDT. We observe three main phases: a \textit{branched polymer-type phase} (phase A), a \textit{crumpled phase} (phase B) and the physically interesting \textit{de Sitter phase} (phase C). The thicker transition lines represent previously measured phase transition points and the thinner lines an interpolation. Superimposed on the phase diagram are the 4 locations within phase C at which the spectral dimension is determined in this work, as indicated by the black squares. The arrows indicate the apparent direction of decreasing relative lattice spacing.}
\label{PD}
\end{figure}

The canonical point in the physical de Sitter phase of CDT, namely $\left(\kappa_{0}=2.2,\Delta=0.6\right)$, has previously been shown to exhibit a scale dependent spectral dimension, yielding $D_{S}\left(\sigma\rightarrow\infty\right)=4.02\pm 0.10$, and $D_{S}\left(\sigma\rightarrow 0\right)=1.80\pm 0.25$. With a fit to the functional form 

\begin{equation}
D_{S}\left(\sigma\right)=a-\frac{b}{c+\sigma},
\label{funcform} 
\end{equation}

\noindent giving $a=4.02$, $b=119$ and $c=54$ \cite{Ambjorn:2005db}. As the authors of Ref. \cite{Ambjorn:2005db} correctly claim the short distance spectral dimension is thus consistent with the integer 2. However, the fact that this measurement is for just a single point in the parameter space, coupled with the relatively large statistical error makes definitive conclusions difficult. Since this result has potentially important consequences for the renormalizability of gravity, we revisit this calculation, attempting a more comprehensive study of the spectral dimension in phase C of CDT. 

We calculate the spectral dimension as a function of diffusion time for three different $\kappa_{0}$ values along the $\Delta=0.6$ line, in addition to a fourth point $\kappa_{0}=4.4,\Delta=2.0$, within the physical phase of CDT, as indicated by the black squares in Fig. \ref{PD}. For three of these points, namely $\left(\kappa_{0}=2.2,4.4,\Delta=0.6\right)$ and $\left(\kappa_{0}=4.4,\Delta=2.0\right)$, we have also calculated $D_{S}\left(\sigma\right)$ for multiple lattice volumes. This multi-volume study and related discussion can be found in the subsection on systematic errors ( Sec. \ref{systematics}).

In the calculation of the spectral dimension presented in this work we take the starting point of our diffusion to be in the time slice containing the maximal number of $N_{4,1}$ simplices, as is done in e.g. Ref. \cite{Ambjorn:2005db}. In this way we can be sure that we are investigating the bulk properties of the geometry with each diffusion. The diffusion process is followed out to a maximum of 500 diffusion steps. Simulations were performed with a time extension of $t=80$. The attempted Monte Carlo moves that update the geometry were performed in units of $10^{6}$, with each unit defining a sweep. The number of sweeps required to reach a thermalized configuration grows approximately linearly with $N_{4,1}$, and is typically of the order $\sim10^{8}$ sweeps for the largest ensembles \cite{Ambjorn:2005db}. We implement an effective linear four-volume fixing constraint

\begin{equation}
\delta S=\epsilon|N_{4,1}-N^{target}_{4,1}|,
\end{equation}

\noindent with $\epsilon=0.05$ during thermalization and $\epsilon=0.02$ afterwards. We choose to fix $N_{4,1}$ as opposed to the total four-volume $N_{4,1}+N_{3,2}$ for technical convenience. We have checked that for a given number of $N_{4,1}$ simplices we also obtain a sharply peaked number of $N_{3,2}$ simplices, and hence a well-defined average total four-volume $\langle N_{4,1} + N_{3,2} \rangle$ at each point sampled in phase C of the parameter space (see Table \ref{VolComparison}). 

\begin{table}[H]
\centering
\begin{tabular}{|c|c|c|}
\hline
$(\kappa_{0},\Delta)$ & $N_{4,1}$ & $\langle N_{4,1} + N_{3,2} \rangle$ \\ \hline\hline
$\left(2.2,0.6\right)$ & 160,000 & 367,000 \\ \hline
$\left(3.6,0.6\right)$ & 160,000 & 267,000 \\ \hline
$\left(4.4,0.6\right)$ & 160,000 & 207,000 \\ \hline
$\left(4.4,2.0\right)$ & 300,000 & 384,000 \\ \hline
\end{tabular}
\caption{\small A table comparing the number of $N_{4,1}$ simplices with the average total number of simplices $\langle N_{4,1} + N_{3,2} \rangle$ for each point studied in the parameter space.}
\label{VolComparison}
\end{table}

The main results of this work are presented in Fig. \ref{spec1} and Tab. \ref{BPTable}. We find that the long distance spectral dimension is consistent with the semiclassical dimensionality of 4, and that the spectral dimension smoothly decreases to a value consistent with $3/2$ on short distance scales and for sufficiently fine lattice spacings. 

\begin{figure}[H]
  \centering
  \includegraphics[width=1.0\linewidth,natwidth=610,natheight=642]{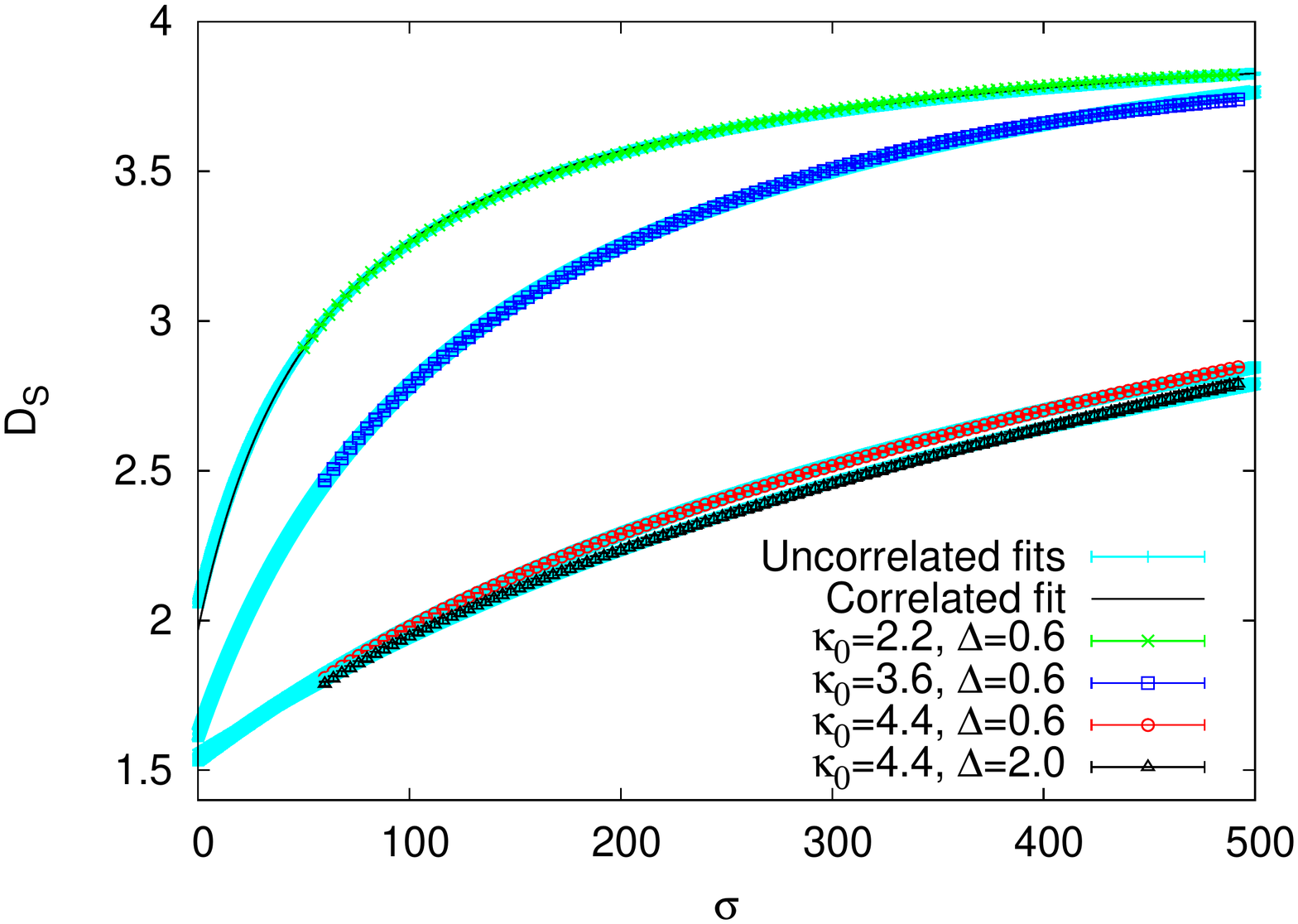}
  \caption{\small The spectral dimension $D_{S}$ as a function of the diffusion time $\sigma$ for four different points in the de Sitter phase of CDT. The $D_{S}\left(\sigma\right)$ curves corresponding to points along the $\Delta=0.6$ line are calculated using 160,000 $N_{4,1}$ simplices. $D_{S}\left(\sigma\right)$ for $\kappa_{0}=4.4,\Delta=2.0$ is calculated using 300,000 $N_{4,1}$ simplices. The light blue error bands come from uncorrelated fits to the data using the functional form of Eq. \ref{funcform} and using the fit range $\sigma \in [50,494]$ for the point $\kappa_{0}=2.2,\Delta=0.6$ and $\sigma \in [60,492]$ for the other three points. We extrapolate to $\sigma=0$ and $\sigma\rightarrow\infty$ using the fit function of Eq. \ref{funcform}. The uncorrelated fit shows only the central value for comparison. Errors presented here are statistical only. Errors in Tab. \ref{BPTable} include the total statistical and systematic error estimate.}
  \label{spec1}
\end{figure}

\begin{table}[H]
\centering                                                                                                        
\begin{tabular}{|c|c|c|c|c|c|}                                                                                     
\hline     
$(\kappa_{0},\Delta)$ & $N_{4,1}$ & $D_{S}(\infty)$ & $D_{S}(0)$ & s.d. of $D_{S}(0)$ from 2 & $a_{rel}$ \\ \hline\hline      
$\left(2.2,0.6\right)$ & 160,000 & $4.05\pm0.17$ & $1.970\pm0.266$ & 0.1 & 1.00 \\ \hline
$\left(3.6,0.6\right)$ & 160,000 & $4.31\pm0.32$ & $1.576\pm0.093$ & 4.5 & 0.57 \\ \hline
$\left(4.4,0.6\right)$ & 160,000 & $4.12\pm0.16$ & $1.534\pm0.058$ & 8.0 & 0.11 \\ \hline
$\left(4.4,2.0\right)$ & 300,000 & $4.14\pm0.12$ & $1.540\pm0.060$ & 7.7 & 0.10 \\ \hline

\end{tabular}
\caption{\small A table of the long $D_{S}(\sigma\rightarrow\infty)$ and short distance spectral dimension $D_{S}(\sigma\rightarrow 0)$ for several different $\left(\kappa_{2},\Delta\right)$ values. $D_{S}(\sigma\rightarrow\infty)$ and $D_{S}(\sigma\rightarrow 0)$ are determined from a fit-function of the form $a-\frac{b}{c+\sigma}$ as first proposed in Ref. \cite{Ambjorn:2005db}. The fifth column gives the number of standard deviations (s.d.) of the values of $D_{S}(\sigma\rightarrow 0)$ from the integer 2. The rescaling factor $a_{rel}$ is determined by the method of best overlap of the rescaled spectral dimension curves.}
\label{BPTable}                                                          
\end{table}

Both correlated and uncorrelated fits to the data give similar results, as demonstrated by the fits to the $\left(2.2,0.6\right)160K$ data in Fig. \ref{spec1}. However, using the full covariance matrix in the estimation of $\chi^{2}$ we obtain a relatively large $\chi^{2}/d.o.f=1.92$. In the absence of any better theoretical guidance as to the correct functional form of the spectral dimension we use an uncorrelated version of the fit function of Eq. \ref{funcform} as our fit ansatz, and attempt to more accurately estimate systematic errors by varying the fit functions and the fit range. We obtain the central values of $D_{S}\left(\infty\right)$ and $D_{S}\left(0\right)$ quoted in Tab. \ref{BPTable} by using the uncorrelated fit function of Eq. \ref{funcform} over the data range $\sigma$ $\in[50,490]$ in steps of 4 for the point $\left(2.2,0.6\right)$ and $\sigma$ $\in[60,490]$ in steps of 4 for the other three points. The errors quoted in Tab. \ref{BPTable} are determined by varying the fit function and the fit range as discussed above and adding the statistical error in quadrature. 

We now return to the holographic argument against the asymptotic safety scenario presented in the introduction. We wish to highlight the fact that Eq's. (\ref{eq:Entropy2}) and (\ref{eq:Entropy3}) agree if, and only if, the spacetime dimension $d$ is equal to 3/2 \interfootnotelinepenalty=10000 \footnote{\scriptsize This counter-argument relies on the plausible assumption that the relevant dimension in the holographic scaling argument is also the spectral dimension as suggested by e.g. Ref. \cite{Carlip:2011uc}}; which is precisely the value we find for the small distance spectral dimension of CDT.\interfootnotelinepenalty=10000 \footnote{\scriptsize In Ref. \cite{Falls:2012nd} the authors argue that the scaling relation of Eq. (\ref{eq:Entropy3}) is incorrect for the class of black hole considered, due to the fact that $R$ depends on the energy $E$ of the black hole, whereas to obtain Eq. (\ref{eq:Entropy3}) $R$ must be treated as a constant. This leads to a modified version of Eq. (\ref{eq:Entropy3}) of the form $\frac{S}{R^{d-1}}\sim \left(\frac{E}{R^{d-1}}\right)^{\nu}$, with $\nu_{cft}=\frac{d-1}{d}$ for a conformal field theory, and $\nu_{BH}=\frac{1}{2}$ for a semiclassical black hole. The authors of Ref. \cite{Falls:2012nd} then point out that $\nu_{BH}=\nu_{CFT}$ when $d=2$ \cite{Falls:2012nd}. This result appears to have some tension with the values we obtain for $D_{S}\left(0\right)$ in this work, at least for some of the points we sampled in phase C of CDT.} The idea that the value of the short distance spectral dimension might resolve the tension between asymptotic safety and holography was first proposed in the context of Euclidean dynamical triangulations \cite{Laihobb}. However, a detailed study of the particular region of parameter space considered the best candidate for a semiclassical phase revealed an effective dimension inconsistent with four dimensional semiclassical spacetime on macroscopic scales \cite{Ambjorn:2013eha,Coumbe:2014nea}. A central motivation of the present work was then to measure the small distance spectral dimension using the causal version of dynamical triangulations (CDT); a formulation known to have a semiclassical phase \cite{Ambjorn:2008wc}.    

\begin{subsection}{Searching for a continuum limit in CDT}

In a lattice formulation of an asymptotically safe field theory, the fixed point would appear as a second-order critical point, the approach to which would define a continuum limit. The divergent correlation length characteristic of a second-order phase transition would allow one to take the lattice spacing to zero while keeping observable quantities fixed in physical units. Hence, developing a method to determine the lattice spacing may prove useful when investigating renormalization group flow within the physical de Sitter phase of CDT, and in particular in the search for a fixed point at which $a\rightarrow 0$. Here we outline one such method that could be used to determine the relative lattice spacing via a comparison of the running spectral dimension at different values of the bare parameters.

Moving along the black line in the direction of the arrows in Fig. \ref{PD} the spectral dimension curves flatten out, as shown in Fig. \ref{spec1}. The implication being that as one increases $\kappa_{0}$ and $\Delta$ the lattice spacing $a$ decreases (similar results were reported in Ref. \cite{Ambjorn:2014gsa}), since it takes a greater number of diffusion steps before the same dimension is obtained. One can then rescale the diffusion time $\sigma$ by a factor $a_{rel}$ for each curve until they overlap, as shown in Fig. \ref{spec2}. Equation (\ref{diffusion1}) seems to suggest the rescaling factor $a_{rel}$ should be proportional to the square of the lattice spacing $a$, which should be taken into account when determining the cut-off scale in physical units. The fit curves are used in the comparison rather than the actual data because it is easier to determine the rescaling factor $a_{rel}$ for which the best overlap occurs. The curves are normalized such that the scale factor $a_{rel}$ is set to unity for the $\kappa_{0}=2.2,\Delta=0.6$ curve.\interfootnotelinepenalty=10000 \footnote{\scriptsize This is obviously a matter of preference and one is free to make any of the $\left(\kappa_{0},\Delta\right)$ points the canonical value against which the others are compared.} The factor $a_{rel}$ that each curve must be rescaled by to obtain agreement with the other curves will then be related to the change in lattice spacing. The rescaling factor $a_{rel}$, as well as the long and short distance spectral dimension, are displayed in Tab. \ref{BPTable} for each $\left(\kappa_{0},\Delta\right)$ value. Interestingly, going from the point $\left(\kappa_{0}=2.2,\Delta=0.6\right)$ to $\left(\kappa_{0}=3.6,\Delta=0.6\right)$ we find qualitatively similar behaviour to that observed in Ref. \cite{Ambjorn:2008wc} between the same two points in parameter space, although the exact quantitative agreement strongly depends on the arguments used. If we assume that the change in the rescaling parameter $a_{rel}$ between different points in the parameter space is proportional to the change in the square of the lattice spacing $a$, as suggested by Eq. (\ref{diffusion1}), and by using the values of the absolute lattice spacing reported in Ref. \cite{Ambjorn:2008wc}, we are led to the conclusion that simulations for the bare parameters $\kappa_{0}=4.4,\Delta=0.6$ and $\kappa_{0}=4.4,\Delta=2.0$ have a lattice spacing already in the sub-Planckian regime.   

\begin{figure}[H]
  \centering
  \includegraphics[width=1.0\linewidth,natwidth=610,natheight=642]{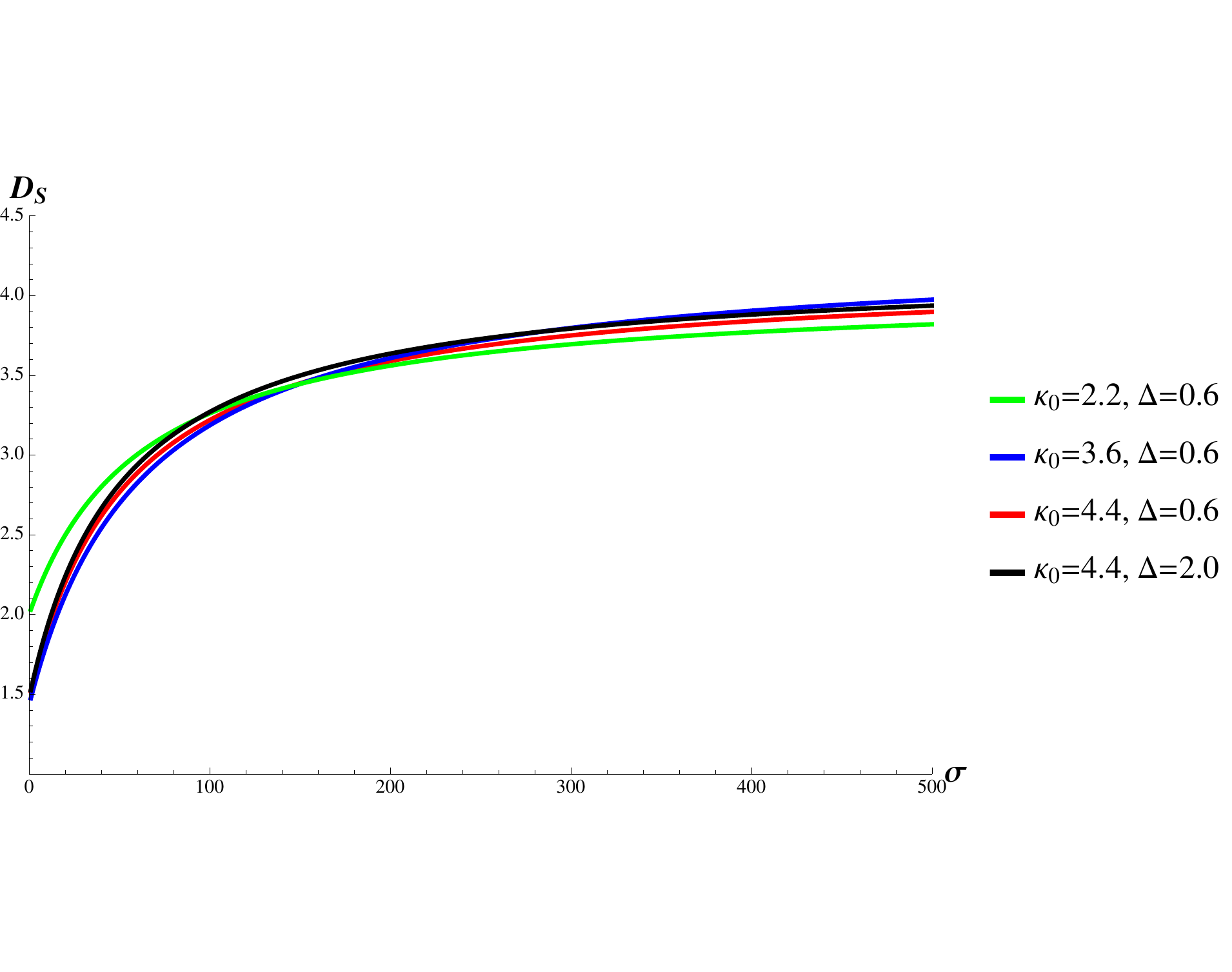}
  \caption{\small Rescaled spectral dimension fits according to the functional form $D_{S}=a-\frac{b}{c+\sigma/a_{rel}}$, with $a_{rel}$ chosen such that the curves give the best overlap.}
  \label{spec2}
\end{figure}

\end{subsection}


\begin{subsection}{Systematic errors}\label{systematics}

Approximating continuous spacetime with a discrete and finite lattice inevitably introduces systematic errors, the main sources being finite-size effects and discretization errors.

Due to finite computational power it is only ever possible to simulate with finite lattice volumes, however, one can quantify finite-size effects by calculating an observable for several different lattice volumes and extrapolating to the infinite volume limit. Thus, one can estimate the lattice volume required such that finite-size effects become negligible. Figures \ref{FSE} and \ref{FSE2} show the spectral dimension as a function of diffusion time for different lattice volumes at three different points in the parameter space. For the point $\kappa_{0}=2.2,\Delta=0.6$ there exists a statistically significant difference between the spectral dimension curves for the 80K and 160K ensembles for large diffusion times. As mentioned in the introduction this is because when $\sigma$ becomes much greater than $N_{4}^{2/D_{S}}$ finite-size effects begin to dominate, eventually driving $D_{S}$ to zero. Finite-size effects can be seen to play a significant role for the 80K ensemble at $\kappa_{0}=2.2,\Delta=0.6$ for $\sigma$ greater than approximately 350, as evidenced by $D_{S}\left(\sigma\right)$ reaching a maximum and then beginning to decrease. However, this is not true of the 120K and 160K ensembles as the condition $\sigma \gg N_{4}^{2/D_{S}}$ is not met for these larger lattice volumes within the $\sigma$ range presented. Furthermore, as we move to points in the parameter space corresponding to finner lattice spacings, i.e. $\kappa_{0}>2.2$ with fixed $\Delta=0.6$, the value of $D_{S}$ is smaller for an equivalent $\sigma$ value, and thus the condition $\sigma \gg N_{4}^{2/D_{S}}$ is only met for much larger $\sigma$ values. 

For the point $\kappa_{0}=4.4,\Delta=2.0$ the much finner lattice spacing results in a much smaller absolute lattice volume, and so one should be careful to simulate with a large enough volume so as to not underestimate the large distance spectral dimension, as suggested in Fig. \ref{FSE2}. Figure \ref{FSE2} indicates that the value of $D_{S}\left(\infty\right)$ increases quite rapidly when comparing the relatively smaller lattice volumes of 160K, 240K and 270K at this point, but that when comparing the larger 270K and 300K ensembles the $D_{S}\left(\sigma\right)$ curves appear to stop growing, becoming statistically comparable. Figures \ref{FSE} and \ref{FSE2} suggest that finite-size effects are mostly under control for the largest lattice volumes at each point, as presented in Fig. \ref{spec1} and Tab. \ref{BPTable}. 

\begin{figure}[H]
  \centering
  \includegraphics[width=1.0\linewidth,natwidth=610,natheight=642]{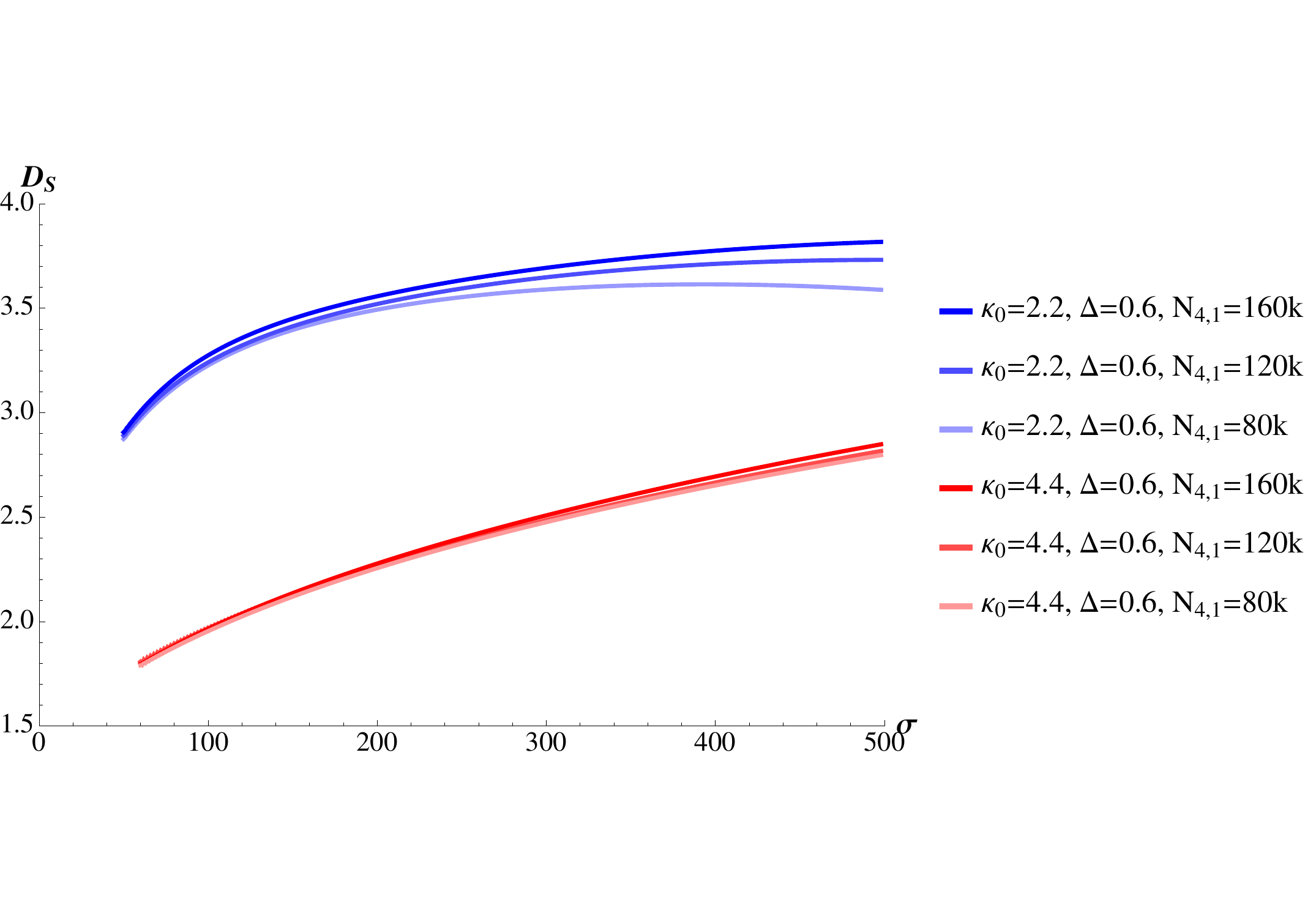}
  \caption{\small A multi-volume study of the spectral dimension at two different points in the parameter space of CDT. Finite-size effects for the $\kappa_{0}=2.2$ and $\kappa_{0}=4.4$ at $\Delta=0.6$ ensembles appear to be under control for the larger 160K lattices.}
\label{FSE}
\end{figure}

\begin{figure}[H]
  \centering
  \includegraphics[width=1.0\linewidth,natwidth=610,natheight=642]{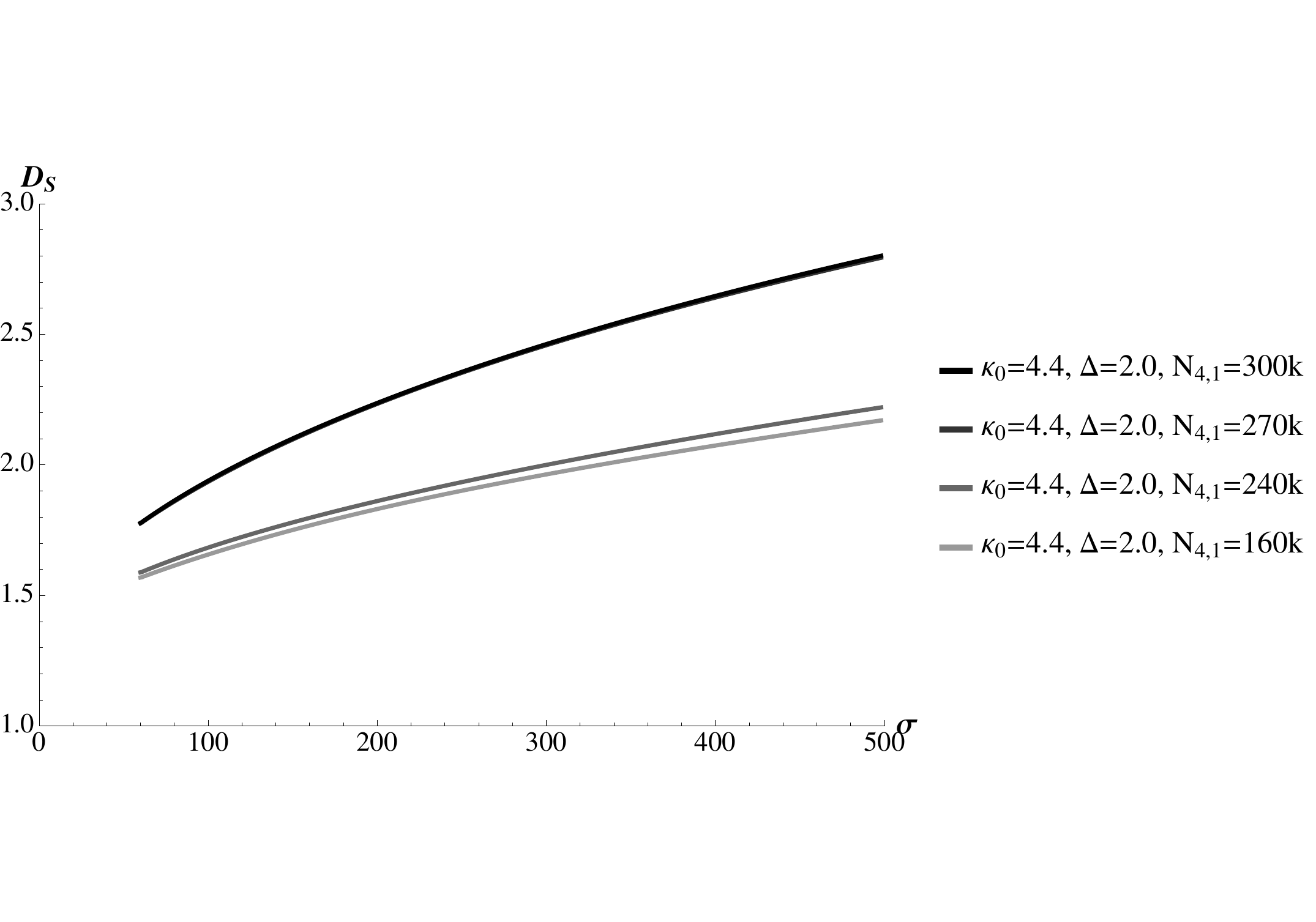}
  \caption{\small A multi-volume study of the spectral dimension at the point $\kappa_{0}=4.4,\Delta=2.0$. Since this point in the parameter space corresponds to a very small lattice spacing in Planck units one must use a much larger lattice volume of 270K or 300K so as to not underestimate the large distance spectral dimension due to the much smaller absolute lattice volume for a given number of $N_{4,1}$ simplices.}
\label{FSE2}
\end{figure}

Errors associated with using a discrete lattice to approximate continuum physics, discretization errors, can be estimated by using an effective field theory and extrapolating down to the continuum. One estimates discretization errors by performing numerical simulations at successively smaller values of the lattice cut-off $a$, i.e. taking the limit $a\rightarrow 0$. Hence, discretization errors become increasingly insignificant as one decreases the lattice spacing. Large discretization errors are typically associated with the small scale spectral dimension. For a small number of diffusion steps the behaviour of $D_{S}\left(\sigma\right)$ can be significantly different when considering an even or odd number of diffusion steps. These odd-even oscillations become negligible for $\sigma\sim 50$ for the coarsest lattice, namely $\kappa_{0}=2.2,\Delta=0.6$, and for $\sigma\sim 60$ for the finer lattices $\kappa_{0}=3.6,4.4,\Delta=0.6$ and $\kappa_{0}=4.4,\Delta=2.0$ as demonstrated in Fig. \ref{OddEven}. To reduce discretization errors we omit values of $D_{S}$ for which $\sigma\leq 50$ for the coarse lattice, and $\sigma\leq 60$ for the finner lattices, from the fit to the functional form of Eq. \ref{funcform}. 

\begin{figure}[H]
  \centering
  \includegraphics[width=1.0\linewidth,natwidth=610,natheight=642]{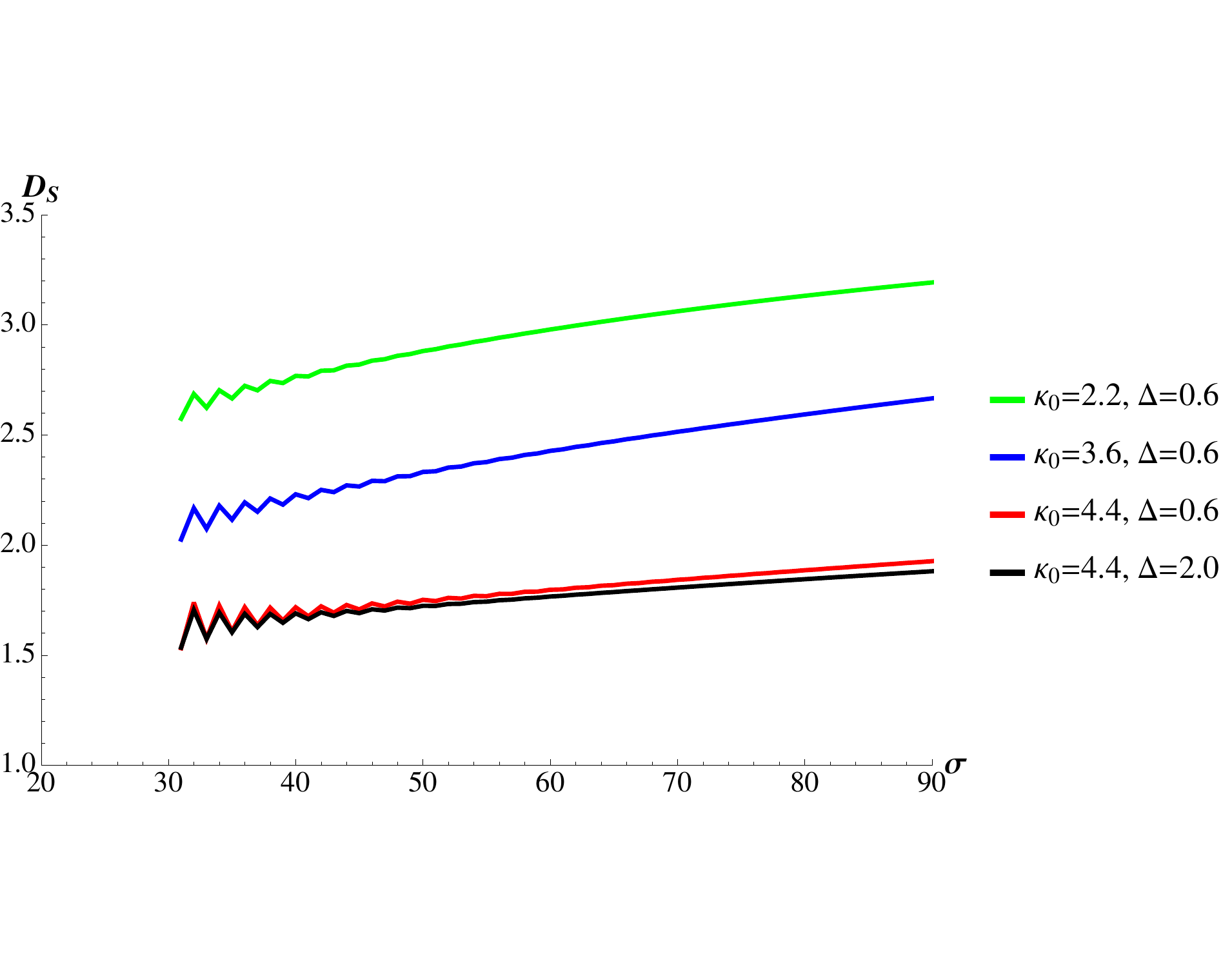}
  \caption{\small Odd-even oscillations in the small scale spectral dimension for four different values of the parameters $\kappa_{0}$ and $\Delta$. Note the oscillations have a larger amplitude and $\sigma$ extension for values of the bare parameters that correspond to finner lattices. In the calculation of $D_{S}\left(\sigma\right)$ we omit $\sigma <50$ values for the coarsest lattice, namely $\kappa_{0}=2.2,\Delta=0.6$, and omit $\sigma <60$ values for the finner lattices $\kappa_{0}=3.6,4.4,\Delta=0.6$ and $\kappa_{0}=4.4,\Delta=2.0$.  
}
\label{OddEven}
\end{figure}

We obtain a more complete estimation of the systematic error associated with our spectral dimension measurements by varying the range of $\sigma$ values over which the fit function of Eq. \ref{funcform} is applied. Furthermore, due to the absence of any solid theoretical motivation for using the functional form of Eq. \ref{funcform} in the fit to our data we also estimate a contribution to the systematic error associated with using the alternative asymptotic functional forms $D_{S}\left(\sigma\right)=a-b\exp{\left(-c\sigma\right)}$ and $D_{S}\left(\sigma\right)=a-\left(b/\left(c+\sigma\right)\right)^{d}$, where a, b, c and d are unconstrained fit parameters, the values of which are given in Tables \ref{AltFits1} and \ref{AltFits2}.

\begin{table}[H]
\centering
\begin{tabular}{|c|cccc|cccc|}
    \hline
Fit-function & \multicolumn{4}{c}{$\left(2.2,0.6\right)$} \vline & \multicolumn{4}{c}{$\left(3.6,0.6\right)$} \vline \\ \hline\hline
   & a&b&c&d & a&b&c&d \\ \hline
$a-b\exp{\left(-c\sigma\right)}$ & 3.74&1.73&0.013&- & 3.74&2.14&0.0078&- \\ \hline
$a-\left(b/\left(c+\sigma\right)\right)^{d}$ & 4.20&108.17&21.69&0.62 & 4.01&479.57&339.14&2.43 \\ \hline
\end{tabular}
\caption{\small The fit parameters a, b, c and d for the two alternative fit functions used in estimating the systematic error for the bare parameters $\left(2.2,0.6\right)$ and $\left(3.6,0.6\right)$ with $N_{4,1}=160,000$.}
\label{AltFits1}
\end{table}

\begin{table}[H]
\centering
\begin{tabular}{|c|cccc|cccc|}
    \hline
Fit-function & \multicolumn{4}{c}{$\left(4.4,0.6\right)$} \vline & \multicolumn{4}{c}{$\left(4.4,2.0\right)$} \vline \\ \hline\hline
   & a&b&c&d & a&b&c&d \\ \hline
$a-b\exp{\left(-c\sigma\right)}$ & 3.83&2.18&0.0016&- & 4.12&2.46&0.0013&- \\ \hline
$a-\left(b/\left(c+\sigma\right)\right)^{d}$ & 3.99&1213.56&586.11&1.24 & 4.00&1337.74&648.10&1.25 \\ \hline
\end{tabular}
\caption{\small The fit parameters a, b, c and d for the two alternative fit functions used in estimating the systematic error for the bare parameters $\left(4.4,0.6\right)$ with $N_{4,1}=160,000$ simplices, and for the bare parameters $\left(4.4,2.0\right)$ with $N_{4,1}=300,000$.}
\label{AltFits2}
\end{table}

\begin{subsubsection}{Investigating systematic errors in phase A}

Using two-dimensional toy models the spectral dimension in the branched polymer phase of Euclidean quantum gravity has been determined from purely analytic considerations \cite{Jonsson:1997gk,Ambjorn:1997jf} to be 4/3. Although such a result is yet to be established in the full four-dimensional theory the geometric properties in phase A of four-dimensional CDT are largely expected to be analogous to the branched polymer phase of EDT. In this work we numerically determine the spectral dimension in phase A of CDT and find a value consistent with the constant 4/3 over the $\sigma$ range studied $\sigma \in [60,492]$, as can be seen in Fig. \ref{BPphaseCDT}. This result suggests that the geometry in phase A of CDT at least shares some universal properties with branched polymer systems. If we assume that the analytical value of 4/3 found in two-dimensional models \cite{Jonsson:1997gk,Ambjorn:1997jf} is also valid in the full four-dimensional theory then it would be possible to get a sense of how small we can reliably take $\sigma$ by comparing our numerical results for the spectral dimension in phase A with the constant value 4/3. Such a comparison also suggests discretization effects are small for $\sigma>60$ for this lattice volume. 

\begin{figure}[H]
  \centering
  \includegraphics[width=1.0\linewidth,natwidth=610,natheight=642]{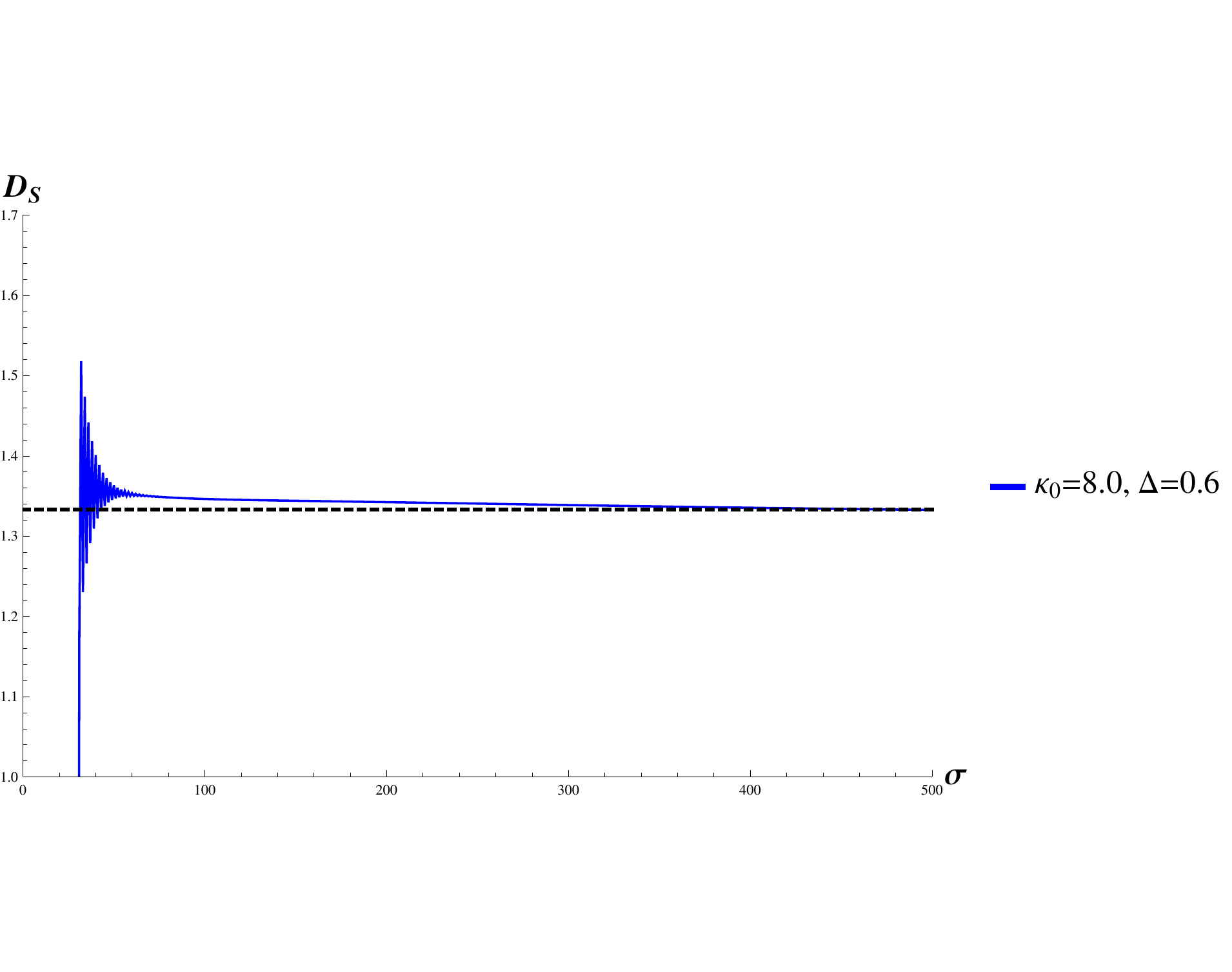}
  \caption{\small The spectral dimension in phase A of CDT, calculated at the point $\kappa_{0}=8.0,\Delta=0.6$ using 160,000 $N_{4,1}$ simplices.}
\label{BPphaseCDT}
\end{figure}

\end{subsubsection}

\end{subsection}

\begin{subsection}{Statistical errors}

If one calculates an observable using a lattice that is not thermalized one will obtain an erroneous result. It is therefore important to check all lattices are thermalized before one begins taking measurements. Once thermalization has been achieved, increasing the number of configurations used in the calculation of the observable will just result in the mean approaching the correct value with an increasingly small statistical error. 

For each point in the parameter space we check that the ensemble is thermalized using two methods. Firstly, we begin with a thermalized smaller volume and allow it to evolve towards a larger target volume. During thermalization, the width of the distribution of $N_{4,1}$ simplices increases very slowly, eventually reaching a plateau. This is the same method of defining thermalization as defined in Ref. \cite{Ambjorn05}. Secondly, after the ensemble has reached a configuration that satisfies the above condition we then plot the observable to be measured as a function of Monte Carlo time and check that there is statistical agreement between the first and second half of the data set over which we perform the measurement.    

Here we apply a best fit to the spectral dimension data using the functional form of Eq. \ref{funcform} and extract values for $D_{S}\left(\sigma\rightarrow\infty\right)$ and $D_{S}\left(\sigma\rightarrow 0 \right)$, plotting them as a function of Monte Carlo time. We conclude that a particular ensemble of triangulations is thermalized over a specific $\sigma$ range if there exists no statistically significant difference between the first and second half of the data range, after passing the first thermalization test. As an example, Figs. \ref{ThermLarge} and \ref{ThermSmall} show the values of $D_{S}\left(\infty\right)$ and $D_{S}\left(0\right)$ for the point $\kappa_{0}=4.4,\Delta=0.6$ as a function of Monte Carlo time using $N_{4,1}=160,000$ simplices. For a configuration number greater than $\sim20,000$ there is no statistical difference in the mean values of $D_{S}\left(0\right)$ and $D_{S}\left(\infty\right)$ when comparing the first and second half of the data set, and we thus conclude this ensemble is thermalized for such a configuration range. All results presented in this work are calculated using thermalized lattices as detailed above. Statistical errors are estimated using a single-elimination jackknife procedure. The total error estimate of our spectral dimension measurements are determined by adding the total systematic and statistical errors in quadrature, and are presented in Tab. \ref{BPTable}.

\begin{figure}[H]
  \centering
  \subfloat{\includegraphics[width=0.85\linewidth,natwidth=610,natheight=642]{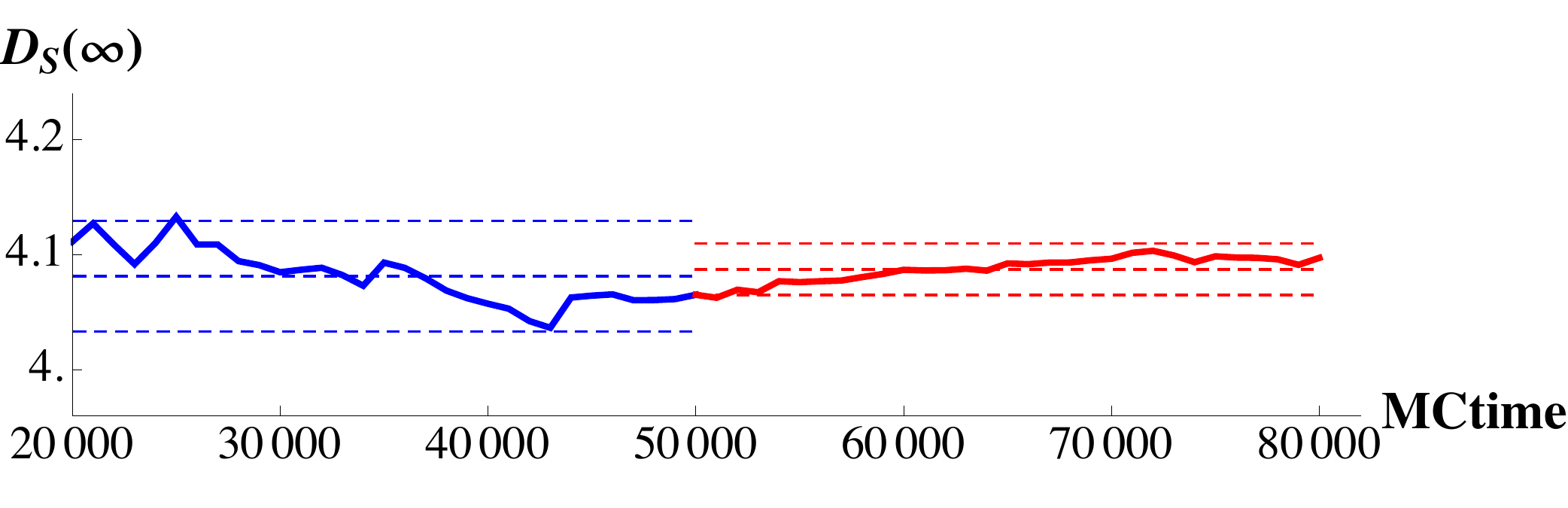}\label{ThermLarge}}\\
\subfloat{\includegraphics[width=0.85\linewidth,natwidth=610,natheight=642]{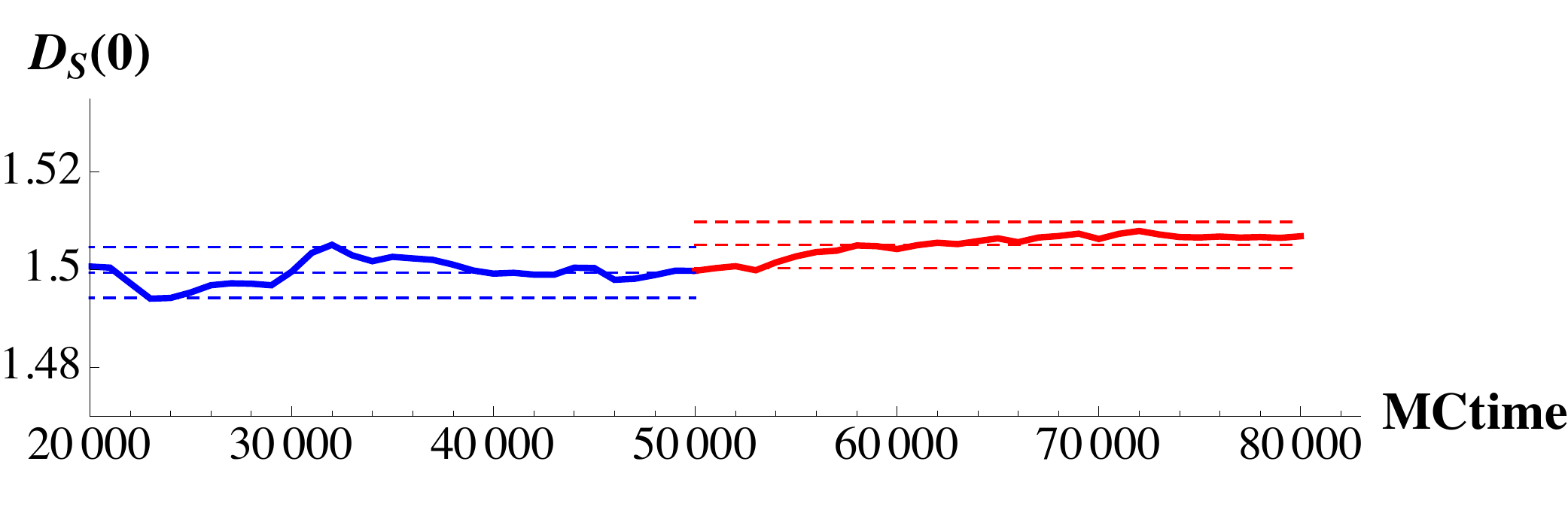}\label{ThermSmall}}
\caption{\small $D_{S}\left(\infty\right)$ and $D_{S}\left(0\right)$ as a function of Monte Carlo time for the bare parameters $\kappa_0=4.4$ and $\Delta=0.6$. The data range we believe to be thermalized is divided into two data sets that are compared with each other for statistical agreement to within 2 standard deviations. The fit function and fit range used to obtain these results are the same as those used in Fig. \ref{spec1}, namely Eq. \ref{funcform} and $\sigma \in [60,492]$, respectively.}
\end{figure}

\end{subsection}

\end{section}


\begin{section}{Discussion and Conclusions}

The aim of this work is to make a more detailed study of dimensional reduction previously found in the CDT approach to quantum gravity, in which a dimensional reduction from $4.02\pm0.1$ on large distance scales, to $1.80\pm0.25$ on small distance scales is reported \cite{Ambjorn:2005db}. The small distance spectral dimension is of particular interest, as a more precise determination of this result could have important implications for the renormalizability of gravity. In this work we give a more detailed study of the running spectral dimension by calculating its value at several different values of the bare parameters and for multiple lattice volumes. Our results are summarised in Tab. \ref{BPTable}. From these results we conclude that the small distance spectral dimension in the de Sitter phase of CDT is more consistent with $3/2$ than with the integer 2, as previously thought \cite{Ambjorn:2005db}. This is the principal result of this work. We wish to point out that this value of the dimension is precisely the value for which Eq's. (\ref{eq:Entropy2}) and (\ref{eq:Entropy3}) agree, and thus it may resolve the tension between asymptotic safety and holography, as originally proposed in Ref. \cite{Laihobb}.

Our studies indicate that as one increases $\kappa_{0}$ and $\Delta$ within the physical phase of CDT the spectral dimension curves flatten out. The implication being that as one moves along such trajectories in the parameter space the lattice spacing $a$ decreases, because for larger values of the bare couplings it takes a greater number of diffusion steps before the same dimension is obtained. One can then rescale the diffusion time by a factor that is related to the relative lattice spacing for each curve until the variance is minimised, i.e. until the curves give ``the best overlap''. This method for determining the relative lattice spacing may prove useful when studying the renormalization group flow in CDT (e.g. Ref. \cite{Ambjorn:2014gsa}), and aid in the search for a putative second order critical point at which one may take a continuum limit.   

The most rapid decrease in the rescaling factor $a_{rel}$ appears to result from maximising $\kappa_{0}$ within phase C of the CDT phase diagram (see Fig. \ref{PD}), and thus tuning $\kappa_{0}$ to its critical value at the first-order transition dividing phase C and phase A. There also seems to be a significantly weaker dependence of $a_{rel}$ on $\Delta$, with $a_{rel}$ appearing to decrease slightly as $\Delta$ increases. Tuning $\kappa_{0}$ to the C-A transition and then studying the effect of varying $\Delta$ on $a_{rel}$ would be a natural next step.

The novel value of the short distance spectral dimension of CDT obtained in this work, $D_{S}\left( 0 \right)\sim 3/2$, differs from the value of $D_{S}\left( 0 \right)\sim 2$ inferred by previous measurements of the spectral dimension of CDT \cite{Ambjorn:2005db}. We find a tension of $\sim8$ standard deviations with the integer value 2 for our finest lattices. Furthermore, the fact that the measurements of $D_{S}\left(\sigma\right)$ presented in this work exhibit a monotonic decrease to a value that is consistent with 3/2, and that $D_{S}\left(0\right)$ shows no sign of changing even for points in the parameter space that appear to be probing the sub-Planckian regime, suggests that our results, at least at present, have some tension with renormalization group predictions that $D_{S}\left( 0 \right)=2$. In light of such suggestive comparisons it may be worth revisiting the renormalization group arguments leading to the result $D_{S}\left(\sigma\rightarrow 0 \right)=2$.

Determining the absolute lattice spacing by measuring fluctuations about de Sitter space, as presented in Ref. \cite{Ambjorn:2008wc}, for all values of the bare couplings investigated in this work would allow one to more thoroughly assess the reliability of using the rescaling of the spectral dimension in determining the change in lattice spacing. Furthermore, determining the absolute lattice spacing via the method presented in Ref. \cite{Ambjorn:2008wc} for points corresponding to our finest lattices would indicate whether measurements at these values of the bare couplings really are probing the sub-Planckian regime, and possibly give a definitive answer as to whether $D_{S}\left(0\right)$ remains consistent with 3/2 as one probes the manifold on yet smaller distance scales, or whether it begins to increase to $D_{S}\left(0\right)=2$ as expected from renormalization group predictions \cite{Reuter:2012id,Rechenberger:2012pm}. Due to the current absence of such further investigations a definitive comparison between renormalization group and CDT predictions of the short distance spectral dimension is currently incomplete. However, this work is in progress.


\end{section}

\section*{Acknowledgments}

D.N.C is especially grateful to J. Laiho for suggestive early discussions on this work. We thank R. Loll for carefully reading a draft version and providing useful comments. We also thank A. Gorlich, J. Gizbert-Studnicki, F. Saueressig, J. Cooperman and J. Mielczarek for useful discussions. D.N.C acknowledges the support of the grant DEC-2012/06/A/ST2/00389 from the National Science Centre Poland.

\begin{section}{References}

\bibliographystyle{unsrt}
\bibliography{Master_cdtspec}

\begin{thebibliography}{10}

\bibitem{Weinberg79}
Steven Weinberg.
\newblock {General Relativity, an Einstein Centenary Survey}.
\newblock 1997.

\bibitem{Ambjorn:2005db}
J.~Ambjorn, J.~Jurkiewicz, and R.~Loll.
\newblock {Spectral dimension of the universe}.
\newblock {\em Phys.Rev.Lett.}, 95:171301, 2005.

\bibitem{Lauscher:2005qz}
O.~Lauscher and M.~Reuter.
\newblock {Fractal spacetime structure in asymptotically safe gravity}.
\newblock {\em JHEP}, 0510:050, 2005.

\bibitem{Horava:2009if}
Petr Ho{\v r}ava.
\newblock {Spectral Dimension of the Universe in Quantum Gravity at a Lifshitz
  Point}.
\newblock {\em Phys.Rev.Lett.}, 102:161301, 2009.

\bibitem{Modesto:2008jz}
Leonardo Modesto.
\newblock {Fractal Structure of Loop Quantum Gravity}.
\newblock {\em Class.Quant.Grav.}, 26:242002, 2009.

\bibitem{Atick:1988si}
Joseph~J. Atick and Edward Witten.
\newblock {The Hagedorn Transition and the Number of Degrees of Freedom of
  String Theory}.
\newblock {\em Nucl.Phys.}, B310:291--334, 1988.

\bibitem{Calcagni:2013eua}
Gianluca Calcagni and Leonardo Modesto.
\newblock {Nonlocality in string theory}.
\newblock {\em J.Phys.}, A47(35):355402, 2014.

\bibitem{Ambjorn:1991pq}
Jan Ambjorn and Jerzy Jurkiewicz.
\newblock {Four-dimensional simplicial quantum gravity}.
\newblock {\em Phys.Lett.}, B278:42--50, 1992.

\bibitem{Catterall:1994pg}
S.~Catterall, John~B. Kogut, and R.~Renken.
\newblock {Phase structure of four-dimensional simplicial quantum gravity}.
\newblock {\em Phys.Lett.}, B328:277--283, 1994.

\bibitem{Bialas:1996wu}
P.~Bialas, Z.~Burda, A.~Krzywicki, and B.~Petersson.
\newblock {Focusing on the fixed point of 4-D simplicial gravity}.
\newblock {\em Nucl.Phys.}, B472:293--308, 1996.

\bibitem{deBakker:1996zx}
Bas~V. de~Bakker.
\newblock {Further evidence that the transition of 4-D dynamical triangulation
  is first order}.
\newblock {\em Phys.Lett.}, B389:238--242, 1996.

\bibitem{Ambjorn:1998xu}
Jan Ambjorn and R.~Loll.
\newblock {Nonperturbative Lorentzian quantum gravity, causality and topology
  change}.
\newblock {\em Nucl.Phys.}, B536:407--434, 1998.

\bibitem{Ambjorn:2007jv}
J.~Ambjorn, A.~Gorlich, J.~Jurkiewicz, and R.~Loll.
\newblock {Planckian Birth of the Quantum de Sitter Universe}.
\newblock {\em Phys.Rev.Lett.}, 100:091304, 2008.

\bibitem{Ambjorn:2011cg}
J.~Ambjorn, S.~Jordan, J.~Jurkiewicz, and R.~Loll.
\newblock {A Second-order phase transition in CDT}.
\newblock {\em Phys.Rev.Lett.}, 107:211303, 2011.

\bibitem{Ambjorn05}
J.~Ambjorn, J.~Jurkiewicz, and R.~Loll.
\newblock {Reconstructing the universe}.
\newblock {\em Phys.Rev.}, D72:064014, 2005.

\bibitem{Hausdorff:1919vt}
F.~Hausdorff.
\newblock {Dimension und äußeres Maß}.
\newblock {\em Mathematische Annalen 79 (1-2)}, page 157–179, 1919.

\bibitem{Benedetti:2009ge}
Dario Benedetti and Joe Henson.
\newblock {Spectral geometry as a probe of quantum spacetime}.
\newblock {\em Phys.Rev.}, D80:124036, 2009.

\bibitem{Banks:2010tj}
Tom Banks.
\newblock {TASI Lectures on Holographic Space-Time, SUSY and Gravitational
  Effective Field Theory}.
\newblock 2010.

\bibitem{Shomer:2007vq}
Assaf Shomer.
\newblock {A Pedagogical explanation for the non-renormalizability of gravity}.
\newblock 2007, arXiv/0709.3555.

\bibitem{Falls:2012nd}
Kevin Falls and Daniel~F. Litim.
\newblock {Black hole thermodynamics under the microscope}.
\newblock 2012.

\bibitem{Percacci:2010af}
Roberto Percacci and Gian~Paolo Vacca.
\newblock {Asymptotic Safety, Emergence and Minimal Length}.
\newblock {\em Class.Quant.Grav.}, 27:245026, 2010.

\bibitem{Koch:2013owa}
Benjamin Koch and Frank Saueressig.
\newblock {Structural aspects of asymptotically safe black holes}.
\newblock {\em Class.Quant.Grav.}, 31:015006, 2014.

\bibitem{Carlip:2011uc}
S.~Carlip and D.~Grumiller.
\newblock {Lower bound on the spectral dimension near a black hole}.
\newblock {\em Phys.Rev.}, D84:084029, 2011.

\bibitem{Laihobb}
J.~Laiho and D.~Coumbe.
\newblock {Evidence for Asymptotic Safety from Lattice Quantum Gravity}.
\newblock {\em Phys.Rev.Lett.}, 107:161301, 2011.

\bibitem{Ambjorn:2013eha}
J.~Ambjorn, L.~Glaser, A.~Goerlich, and J.~Jurkiewicz.
\newblock {Euclidian 4d quantum gravity with a non-trivial measure term}.
\newblock {\em JHEP}, 1310:100, 2013.

\bibitem{Coumbe:2014nea}
Daniel Coumbe and John Laiho.
\newblock {Exploring Euclidean Dynamical Triangulations with a Non-trivial
  Measure Term}.
\newblock 2014.

\bibitem{Ambjorn:2008wc}
J.~Ambjorn, A.~Gorlich, J.~Jurkiewicz, and R.~Loll.
\newblock {The Nonperturbative Quantum de Sitter Universe}.
\newblock {\em Phys.Rev.}, D78:063544, 2008.

\bibitem{Ambjorn:2014gsa}
J.~Ambjorn, A.~Goerlich, J.~Jurkiewicz, A.~Kreienbuehl, and R.~Loll.
\newblock {Renormalization Group Flow in CDT}.
\newblock 2014.

\bibitem{Jonsson:1997gk}
Thordur Jonsson and John~F. Wheater.
\newblock {The Spectral dimension of the branched polymer phase of
  two-dimensional quantum gravity}.
\newblock {\em Nucl.Phys.}, B515:549--574, 1998.

\bibitem{Ambjorn:1997jf}
Jan Ambjorn, Dimitrij Boulatov, Jakob~L. Nielsen, Juri Rolf, and Yoshiyuki
  Watabiki.
\newblock {The Spectral dimension of 2-D quantum gravity}.
\newblock {\em JHEP}, 9802:010, 1998.

\bibitem{Reuter:2012id}
Martin Reuter and Frank Saueressig.
\newblock {Quantum Einstein Gravity}.
\newblock {\em New J.Phys.}, 14:055022, 2012.

\bibitem{Rechenberger:2012pm}
Stefan Rechenberger and Frank Saueressig.
\newblock {The R2 phase-diagram of QEG and its spectral dimension}.
\newblock {\em Phys.Rev.}, D86:024018, 2012.

\end{thebibliography}

\end{section}


\end{document}